\documentclass[conference]{IEEEtran}
\IEEEoverridecommandlockouts
\usepackage{amsmath,amssymb,amsfonts}
\usepackage{graphicx}
\usepackage{textcomp}
\usepackage{xcolor}
\def\BibTeX{{\rm B\kern-.05em{\sc i\kern-.025em b}\kern-.08em    T\kern-.1667em\lower.7ex\hbox{E}\kern-.125emX}}
\usepackage{verbatim}
\usepackage{paralist}
\usepackage{graphics} 
\usepackage{epsfig} 
\usepackage{mathptmx} 
\usepackage{times} 
\usepackage{physics}
\usepackage{mdwmath}
\usepackage{mdwtab}
\usepackage[hidelinks]{hyperref}
\usepackage{algpseudocode}
\usepackage{algorithm}
\usepackage{caption}
\usepackage{subcaption}

\newtheorem{remark}{Remark}
\usepackage{tikz}
\usetikzlibrary{shapes, arrows.meta, positioning}

\begin{document}

\title{
A Hybrid Quantum-Classical Physics-Informed Neural Network Architecture for Solving Quantum Optimal Control Problems
\\
\thanks{}
}

\author{\IEEEauthorblockN{Nahid Binandeh Dehaghani}
\IEEEauthorblockA{
\textit{SYSTEC, ARISE}\\
\textit{Faculty of Engineering,\,University of Porto}\\
Porto, Portugal \\
nahid@fe.up.pt}
\and
\IEEEauthorblockN{A. Pedro Aguiar}
\IEEEauthorblockA{
\textit{SYSTEC, ARISE}\\
\textit{Faculty of Engineering,\,University of Porto}\\
Porto, Portugal \\
pedro.aguiar@fe.up.pt}
\and
\IEEEauthorblockN{Rafal Wisniewski}
\IEEEauthorblockA{
\textit{Department of Electronic Systems} \\ 
\textit{Aalborg University}\\
Aalborg, Denmark \\
raf@es.aau.dk}
}

\maketitle

\begin{abstract}
This paper proposes an integrated quantum-classical approach that merges quantum computational dynamics with classical computing methodologies tailored to address control problems based on Pontryagin's minimum principle within a Physics-Informed Neural Network (PINN) framework.
By leveraging a dynamic quantum circuit that combines Gaussian and non-Gaussian gates, the study showcases an innovative approach to optimizing quantum state manipulations. The proposed hybrid model
effectively applies machine learning techniques to solve 
optimal control problems. 
This is illustrated through the design and implementation of a hybrid PINN network to solve a quantum state
transition problem in a two and three-level system, highlighting its potential across various quantum computing applications.
\end{abstract}


\section{Introduction to Quantum Technologies}
In the realm of quantum control, advances have been made through various methodologies, including controllable dissipative dynamics \cite{jamonneau2016coherent,lin2013dissipative}, the backaction induced by measurement \cite{branczyk2007quantum,montenegro2017macroscopic}, Lyapunov control \cite{dehaghani2024stochastic}, the implementation of optimal control theory \cite{Dehaghani2023QuantumST}, and the utilization of differentiable programming \cite{coopmans2021protocol}. These techniques are instrumental in achieving goals such as the preservation of quantum states \cite[Chapter 12]{cong2014control}, efficient state transitions \cite{Dehaghani2023QuantumST}, dynamical decoupling in dissipative quantum systems \cite{zhen2016optimal,viola2005random}, and precise trajectory tracking \cite{wu2022trajectory}. Remarkably, these approaches have found applications across a diverse array of platforms such as atomic configurations \cite{du2014experimental}, light-matter interactions \cite{tancara2021steering}, solid-state mechanisms \cite{tian2019optimal}, and trapped ion systems \cite{alonso2013quantum}. The essence of dynamical quantum control lies in manipulating the quantum state evolution via time-dependent Hamiltonians \cite{dehaghani2022quantum}, constrained by factors such as laser intensity, frequency broadening, and relaxation phenomena. Examples of constrained quantum optimal control problems can be found in \cite{dehaghani2023quantum,Dehaghani2023QuantumST}.

Quantum optimal control, grounded in optimal control theory \cite{lewis2012optimal,kirk2004optimal,vinter2010optimal}, emerges as a preeminent strategy among quantum control methods (see \cite{dehaghani2022quantum} for details), transforming state manipulation challenges into global optimization problems. This strategy entails identifying a set of permissible controls that adhere to the system's dynamic equations while minimizing a predefined cost function, which could vary based on the specific demands of the quantum control task—such as reducing control duration, energy consumption, or error margins between initial and target states. Techniques from conventional optimal control, including variational methods \cite{berkani2012optimal}, the Pontryagin minimum principle \cite{naidu2018pontryagin}, and iterative algorithms \cite{sidar1968iterative}, have been successfully adapted to quantum contexts. These methodologies have shown considerable utility in manipulating quantum phenomena within fields like physical chemistry and nuclear magnetic resonance studies. More specifically, the application of Pontryagin's optimality conditions to quantum systems has been studied in \cite{dehaghani2022optimal}.

The quest for optimal quantum control sequences is inherently complex and system-specific, necessitating advanced computational strategies for the simulation of quantum systems. Innovations such as neural network-based parametrization have facilitated the simulation of many-body wavefunctions in both closed \cite{carleo2017solving, cai2018approximating,choo2018symmetries} and open (dissipative) quantum systems \cite{luschen2017signatures, daley2014quantum, hartmann2019neural,overbeck2016time}. Furthermore, research has branched into hybrid models \cite{koch2021neural,liu2021hybrid}, probabilistic approaches using positive operator-valued measures \cite{luo2022autoregressive}, and data-driven frameworks \cite{mazza2021machine}, underscoring the complexity of modeling open quantum system dynamics. Machine learning stands out as a flexible and potent tool in expanding our capabilities to address these challenges, although fully integrating quantum control dynamics and artificial intelligence within a unified deep learning solution remains uncharted territory. Neural networks, traditionally reliant on data for training, could benefit from models that inherently comply with physical laws. To this end, Physics-Informed Neural Networks (PINNs) have been introduced, requiring only the system's mathematical model for training \cite{karniadakis2021physics,raissi2019physics}. This approach has shown promise in solving complex problems like high-dimensional partial differential equations \cite{zeng2022adaptive}, many-body quantum systems \cite{norambuena2024physics}, quantum fields \cite{martyn2023variational}, and in extracting physical principles from sparse and noisy data \cite{desai2021port}.

\textbf{Main Contributions: }
In this work, we build upon our established framework of classical PINNs, previously tailored to harness the Pontryagin Minimum Principle (PMP) \cite{johnston2021theory,d2021pontryagin,johnston2020fuel}, specifically for quantum systems control \cite{dehaghani2023quantum,dehaghani2023application}. Our proposed contribution extends this approach into a hybrid quantum-classical domain, innovating beyond traditional models by replacing the traditional neural network model with a dynamic quantum neural network (QNN) circuit comprising Gaussian and non-Gaussian gates, complying with the seminal works in \cite{killoran2019continuous,markidis2022physics}. Our approach aligns with foundational research in continuous variable quantum systems as outlined in significant prior studies \cite{fukui2022building}. 
This architectural advancement not only adheres to but also expands the theoretical and practical applications of QNNs in tackling complex optimal control problems
for optimizing photon number manipulation strategies over time.

We implement a strategic configuration of quantum gates, orchestrated through Strawberry Fields and optimized within a TensorFlow environment. 
These gates are dynamically parameterized, with time-varying parameters fine-tuned to minimize a loss function throughout the process.
These parameters are adjusted through an iterative training process, using 
Adam optimizer, known for its effectiveness in handling large datasets and complex variable spaces.
We employ TensorFlow's GradientTape to dynamically monitor and calculate gradients during the simulation, allowing for efficient backpropagation and optimization of quantum gate parameters. This mechanism is crucial for adapting the quantum gates to the evolving requirements of the control problem. The training regimen consists of repeated simulation epochs, where quantum states are evolved according to the parameterized instructions defined within our quantum physics-informed neural network (QPINN). Each epoch of the training refines the quantum gate parameters based on the observed output and the target control objectives, effectively learning the optimal control strategy over time. This process mimics the feedback mechanism in classical control systems but is executed within a quantum framework.
This integrated approach showcases the significant potential of merging quantum and classical computing to solve intricate quantum control problems.

Our hybrid model's main contribution lies in its capability to seamlessly merge quantum computational dynamics with classical neural network methodologies, thus providing a robust framework showcasing its capability for the simulation and control of open quantum systems. By leveraging the unique properties of quantum mechanics, such as state superposition and entanglement, our system enhances the precision and efficiency of control strategies. These strategies are particularly adept at managing the evolution and stability of quantum states under varying operational conditions.
This approach allows for a versatile manipulation of quantum states, offering a broader range of control dynamics compared to traditional single-model systems. This versatility is crucial for addressing the complex requirements of modern quantum technologies, including quantum computing, secure communications, and precision metrology. Through detailed simulations, we demonstrate the practical effectiveness of our approach, underscoring its potential to significantly impact the field of quantum optimal control by providing new tools and methodologies that are capable of advancing the frontier of quantum technology applications.


\textbf{Paper Structure: }The structure of the work is as follows: In Section II, we explain the developed hybrid quantum-classical optimization approach through two subsections. We first explain the quantum circuit exploited in our method, and then, we delve into our hybrid approach based on the theory of functional connections. In Section III, we explain a quantum optimal control problem. We explain the system under study by introducing the Lindblad master equation and facilitate our computations by mapping the quantum density operator to a real valued state vector. We then obtain the necessary optimality conditions in the form of Pontryagin's Minimum Principle equipped with saturation function and system extension technique. In order to solve the resulting boundary value problem, we use the theory of functional connections, where the approximated solution is obtained by means of a QNN unit, explained in Section II. The simulation results have been shown for the quantum state transition problem in a two-level system and also a three-level system in Section IV. The paper ends with a conclusion and an overview of prospective research challenges. 

\begin{figure*}[t]
  \includegraphics[width=\textwidth]{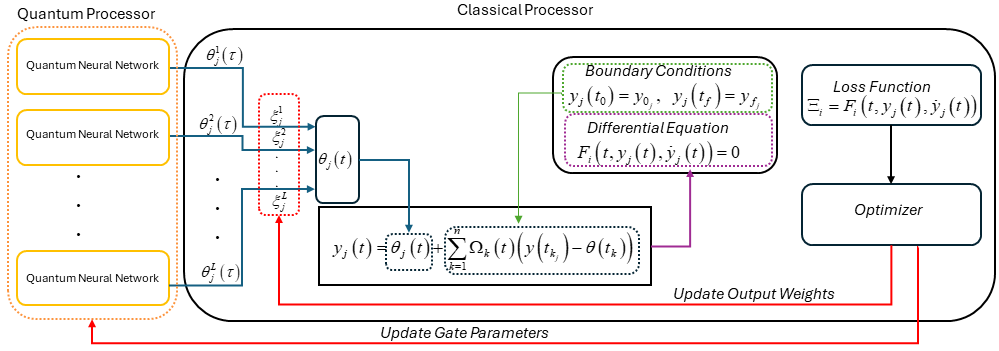}
\caption{An integrated quantum-classical approach to solve differential equations. Here, the subscript $i$ indicates a particular differential equation from a finite set forming the system of differential equations, $j$ is the index of unknown functions of time which are the solutions to the system, and $L$ indicates the number of quantum neural network in the quantum processor. 
The circuit output, symbolized as $\theta_j(t)$, informs the formulation of solution $y_j(t)$, subject to initial and terminal conditions, ensuring boundary conditions adherence and enabling the substitution of the solution into the differential equation framework.
This is followed by the computation of a loss function that quantifies the error in satisfying the differential equation, which an optimizer then minimizes to derive the approximated solution. The optimizer updates both the output weights $\xi_j^l$ and the gate parameters. The optimizer
effectively links the quantum computation outputs with the classical optimization objectives.}
  \label{PIQNN}
\end{figure*}

\section{Hybrid Quantum-Classical Optimization}
The indicated QPINN architecture is built upon a hybrid quantum-classical model, where a surrogate neural network developed in the continuous variable quantum framework approximates the solutions to the differential equations governing the system dynamics. 
Our QPINN model, depicted in Fig. \ref{PIQNN} merges QNN, implemented in a continuous variable quantum framework, with the theory of functional connections.
The core of our implementation involves defining the quantum circuits and the optimization algorithm necessary for training the QPINNs. 
In this section, we first delve into the implementation of the exploited quantum circuit and continuous variable model. In the second subsection, we explain the implementation of our hybrid approach based on the theory of functional connections.

\subsection{Quantum Circuit implementation}
The continuous variable quantum computing framework is centered around the concept of the qumode, which serves as the fundamental unit of information. In this framework, the qumode 
$\left| \psi  \right\rangle $ is represented through the basis expansion of quantum states, as $\left| \psi  \right\rangle =\int{\psi \left( x \right)}\left| x \right\rangle dx$ where the states represent the eigenstates of the $\hat{x}-$quadrature, $\hat{x}\left|x\right\rangle={x}\left|x\right\rangle$, with $x\in \mathbb{R}$ is the eigenvalue. 
The continuous variable quantum computing model presents an alternative to the widely recognized qubit-based approach. In the qubit framework, quantum information is represented using discrete states, as exemplified by a qubit \( |\phi\rangle \) which can be expressed as a linear combination of the basis states \( |0\rangle \) and \( |1\rangle \):
$|\phi\rangle = \phi_0 |0\rangle + \phi_1 |1\rangle$.
Conversely, continuous variable quantum computing utilizes the concept of qumodes, where states are defined across a continuum of possible values, rather than discrete coefficients. This approach employs a continuous spectrum of coefficients, reflecting a continuous eigenvalue spectrum. This framework is particularly suited to scenarios involving continuous quantum operators, aligning seamlessly with the principles of continuous variable quantum computing. Here, we consider each qumode $i$ associated with a pair of non-commuting position $\hat{x}_i$ and momentum $\hat{p}_i$ operators. These operators constitute the phase space and are good instances of continuous quantum variables, defined as
$\hat{x}=\int_{-\infty }^{\infty }{x\left| x \right\rangle }\left\langle  x \right|dx$ and $\hat{p}=\int_{-\infty }^{\infty }{p\left| p \right\rangle }\left\langle  p \right|dp$ with $\left| x \right\rangle$ and $\left| p \right\rangle$ being orthogonal vectors. The non commuting nature of these operators leads to the Heisenberg
uncertainty principle for the simultaneous measurements of $\hat{x}$ and $\hat{p}$.

Similar to the established qubit-based approach, continuous variable quantum computing can be described through the use of basic gates, which may be implemented as optical devices. A continuous variable  quantum program typically involves a sequence of gates affecting one or several qumodes. Essential to the development of continuous variable quantum neural networks are the following four basic Gaussian gates \cite{killoran2019continuous}
\begin{itemize}
    \item \textbf{Displacement Gate} This gate translates the state within the phase space by a complex number \(\alpha\), effectively modifying both the position and momentum coordinates by \(\alpha\)'s real and imaginary components, respectively. Symbolically, the transformation is represented as a shift in the phase space variables.\\
    \begin{tikzpicture}[gate/.style={rectangle, draw, fill=blue!20, text width=3em, text centered, rounded corners, minimum height=2em},
    line/.style={draw, -Latex}]
    
    \node (input1) {\small
    $\left[ \begin{matrix}
   x  \\
   p  \\
\end{matrix} \right]$};
    
    \node [gate, right=0.25cm of input1] (displacement) {\small D-gate: $D(\alpha)$};
    
    \node [right=0.25cm of displacement] (output1) {\small 
    $\left[ \begin{matrix}
   x+{{\alpha}_{\operatorname{Re}}}  \\
   p+{{\alpha}_{\operatorname{Im}}}  \\
\end{matrix} \right]$};

    \draw[line] (input1) -- (displacement);
    \draw[line] (displacement) -- (output1);
\end{tikzpicture}

\item \textbf{Rotation Gate} $R(\varphi)$: This gate rotates the state in phase space around the origin by an angle \(\varphi\), where \(\varphi\) spans the interval \([0, 2\pi]\). The action of this gate can be represented by a rotation matrix that multiplies the phase space vector, leading to a new configuration defined by the rotated coordinates.\\
\begin{tikzpicture}[gate/.style={rectangle, draw, fill=blue!20, text width=3em, text centered, rounded corners, minimum height=2em},
    line/.style={draw, -Latex}]
    
    \node (input2) {\small
    $\left[ \begin{matrix}
   x  \\
   p  \\
\end{matrix} \right]$};
    
    \node [gate, right=0.25cm of input2] (rotation) {\small R-gate: $R(\varphi)$};
    
    \node [right=0.25cm of rotation] (output2) {
    \small$\left[ \begin{matrix}
   \cos \left( \varphi  \right) & \sin \left( \varphi  \right)  \\
   -\sin \left( \varphi  \right) & \cos \left( \varphi  \right)  \\
\end{matrix} \right]\left[\begin{matrix}
   x  \\
   p  \\
\end{matrix} \right]$};

    \draw[line] (input2) -- (rotation);
    \draw[line] (rotation) -- (output2);
\end{tikzpicture}

\item \textbf{Squeezing Gate} $S(r)$: By applying the squeezing gate, the quantum state experiences a scaling transformation that alters the variances of the position and momentum. The scaling factor \(r\), a real number, either compresses or expands the state in the phase space, resulting in a `squeezed' state that has reduced uncertainty in one variable while increasing it in the conjugate variable, in accordance with Heisenberg's uncertainty principle.
\begin{tikzpicture}[gate/.style={rectangle, draw, fill=blue!20, text width=3em, text centered, rounded corners, minimum height=2em},
    line/.style={draw, -Latex}]
    
    \node (input3) {\small$\left[  \begin{matrix}
   x  \\
   p  \\
\end{matrix} \right]$};
    
    \node [gate, right=0.25cm of input3] (squeezing) {\small  S-gate: $S(r)$};
    
    \node [right=0.25cm of squeezing] (output3) {\small$\left[  \begin{matrix} 
   e^{-r} & 0 \\
   0 & e^r \\
\end{matrix} \right] \left[ \begin{matrix}
   x  \\
   p  \\
\end{matrix} \right]$};

    \draw[line] (input3) -- (squeezing);
    \draw[line] (squeezing) -- (output3);
\end{tikzpicture}
  
  \item \textbf{Beam-Splitter Gate} $BS(\theta)$: Acting on two qumodes, the beam-splitter gate mixes their states, analogous to the interference effects observed in classical optics. It is mathematically described by a transformation that combines rotation-like effects on the two qumodes' phase space coordinates, parameterized by an angle \(\theta\) in the range \([0, 2\pi]\). The output state is a hybridization of the input states, with the mixing determined by \(\theta\).
\begin{tikzpicture}[gate/.style={rectangle, draw, fill=blue!20, text width=2em, text centered, rounded corners, minimum height=2em},
    line/.style={draw, -Latex}]
    
    \node (input4) {\!\!\!\!\!\!\!\!\small$\left[ \begin{matrix}
   x_1  \\
   x_2  \\
   p_1  \\
   p_2  \\
\end{matrix} \right]$};
 \!\!\!\!\!\!\!   
    \node [gate, right=0.25cm of input4] (beamSplitter) {\small
    $BS(\theta)$};
    
    \node [right=0.25cm of beamSplitter] (output4) {\small $\left[\begin{matrix}
   \cos(\theta) & -\sin(\theta) & 0 & 0 \\
   \sin(\theta) & \cos(\theta) & 0 & 0 \\
   0 & 0 & \cos(\theta) & -\sin(\theta) \\
   0 & 0 & \sin(\theta) & \cos(\theta) \\
\end{matrix} \right] \left[ \begin{matrix}
   x_1 \\
   x_2 \\
   p_1 \\
   p_2 \\
\end{matrix} \right]$};
    \draw[line] (input4) -- (beamSplitter);
    \draw[line] (beamSplitter) -- (output4);
\end{tikzpicture}  
\end{itemize}
A crucial composite gate, the interferometer, can be constructed using a sequence of beam-splitter and rotation gates. When only one qumode is involved, the function of the interferometer simplifies to that of a rotation gate. These Gaussian gates together facilitate the creation of an affine transformation as implemented in \cite{killoran2019continuous,markidis2022physics}, which plays a pivotal role in enabling the computational processes of neural networks.

Beyond the four primary Gaussian gates outlined earlier, we use a non-Gaussian gate, Kerr gate, introducing a type of non-linearity akin to that of activation functions in classical neural networks. Notably, the integration of non-Gaussian gates with Gaussian gates in a sequence significantly enhances the capabilities of continuous variable quantum circuits. This combination achieves universality, ensuring that any continuous variable state can be generated with at most polynomial computational overhead. 
\begin{itemize}
    \item \textbf{Kerr Gate} $K(\kappa)$:
This gate applies a nonlinear phase shift proportional to the square of the photon number, introducing significant nonlinearity into quantum circuits. The transformation enacted by the Kerr gate \( K(\kappa) \) on a photon number state \( |n\rangle \) can be expressed as \( |n\rangle \rightarrow \exp(i \kappa n^2) |n\rangle \), where \( \kappa \in \mathbb{R} \) serves as the gate parameter, and is the strength of the nonlinearity. This transformation is crucial for tasks that Gaussian gates alone cannot perform, such as specific types of quantum state entanglement and enhanced quantum computation processes.

\begin{tikzpicture}[gate/.style={rectangle, draw, fill=blue!20, text width=3em, text centered, rounded corners, minimum height=2em},
    line/.style={draw, -Latex}]
    
    \node (inputK) {\small
    $\left[ \begin{matrix}
   |n\rangle  \\
\end{matrix} \right]$};
    
    \node [gate, right=0.25cm of inputK] (kerr) {\small K-gate: $K(\kappa)$};
    
    \node [right=0.25cm of kerr] (outputK) {\small 
    $\left[ \begin{matrix}
   \exp(i \kappa n^2) |n\rangle \\
\end{matrix} \right]$};

    \draw[line] (inputK) -- (kerr);
    \draw[line] (kerr) -- (outputK);
\end{tikzpicture}
\end{itemize}

\begin{figure}[t]
\centerline{\includegraphics[scale=0.4]{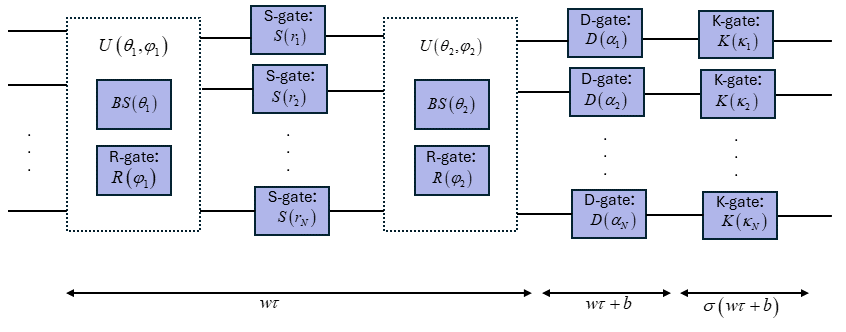}}
\caption{Schematic depiction of a quantum neural network unit utilizing continuous variable quantum gates. The architecture integrates unitary transformations \( U(\theta_i, \phi_i) \), consisting of beam-splitter \( BS(\theta) \) and rotation \( R(\phi) \) gates, along with squeezing operations \( S(r_i) \) and followed by displacement \( D(\alpha_i) \) gates to implement affine transformations, analogous to the neural network's weights and biases. 
This setup portrays the flow from input through quantum processing stages, resulting in a non-linearly transformed output introduced by the Kerr gates \( K(\kappa_i) \), thereby drawing a parallel with classical neural network operations.}
\label{fig:quantum_neural_network}
\end{figure}

\begin{figure}[t]
\centerline{\includegraphics[scale=0.55]{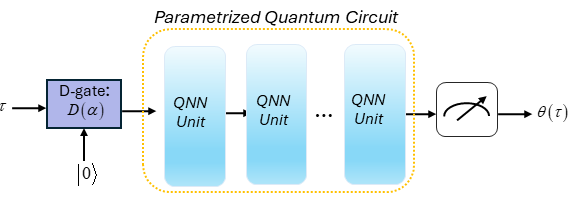}}
\caption{The input $\tau$ is encoded by means of a displacement gate on the vacuum. Then, the approximator network, illustrated by a parametrized quantum circuit consisting of several connected quantum neural network units, whose output $\theta(t)$ is obtained by applying measurement.}
\label{fig:QNN}
\label{QNNU}
\end{figure}

Having defined the basic five continuous variable quantum gates, we now delve into the circuit structure of a quantum neural network unit as depicted in Fig. \ref{fig:quantum_neural_network}. The quantum neural network unit comprise a sequence consisting of a first interferometer, local squeeze gates, a second interferometer, local displacements, and non-Gaussian Kerr gates.
The interferometers are general N-port linear optical elements containing beamsplitter $BS(\theta)$ and rotation $R(\varphi)$ gates.  
The outcome obtained from the combination of the first interferometer $U(\theta_1,\varphi_1)$, local squeeze gates $\otimes _{i=1}^{N}S\left( {{r }_{i}} \right)$ and $\otimes _{i=1}^{N}S\left( {{r}_{i}} \right)$, a second interferometer $U(\theta_2,\varphi_2)$  effectively multiplies the phase space vector by the neural network weights $w$, which are determined by the parameters of the interferometers and squeezing gates. Similar to classical neural networks, the displacement gate $\otimes _{i=1}^{N}D\left( {{\alpha }_{i}} \right)$ serves to add a bias term $b$. The sequence of these Gaussian transformations, i.e., $\otimes _{i=1}^{N}D\left( {{\alpha }_{i}} \right) \circ U(\theta_2,\varphi_2) \circ \otimes _{i=1}^{N}S\left( {{r}_{i}} \right) \circ U(\theta_1,\varphi_1)$ is enough for parametrization of all possible unitary affine transformation on $N$ qumodes.
Ultimately, a Kerr gate introduces non-linearity, analogous to the activation function $\sigma$ used in classical neural networks, therefore, for an input $\tau$ one attains $\left| \tau  \right\rangle \to \left| \sigma \left( w\tau +b \right) \right\rangle$.

By arranging multiple quantum neural units in sequence, a quantum neural network can be constructed, as illustrated in Fig. \ref{fig:QNN}. For a single qumode, control of each gate involves seven gate parameters (\(\alpha\), \(\varphi\), \(r\), \(\theta\), and \(\kappa\)), where \(\varphi\), \(\theta\), and \(\kappa\) are real values, and \(\alpha\) and \(r\) can be either real or complex numbers.
Quantum circuit parameters are categorized into passive and active types. Parameters such as the beam-splitter angles and all gate phases fall into the passive category, whereas parameters for displacement, squeezing, and the magnitude of the Kerr effect are considered active. The objective of training quantum neural networks is to optimize these parameters (\(\alpha\), \(\varphi\), \(r\), \(\theta\), and \(\kappa\)) across various qumodes and units to minimize the hybrid physics-informed neural network loss function.

\subsection{Implementation of a Hybrid Approach Based on the Theory of Functional Connections}
In this section, we combine the illustrated QNN (see Fig. \ref{fig:QNN}) with the theory of functional connection, \cite{johnston2021theory,d2021pontryagin,johnston2020fuel,dehaghani2023quantum,dehaghani2023application} to obtain a hybrid physics-informed neural networks. This theory  
can be used to solve boundary value problems. Consider a generic system composed by $n$ differential equations 
\begin{equation}\label{DE}
F_i\left( t,y(t),\dot y(t)\right)={\bf{0}}, \quad \! \! \!
y_j(t_k)=y_{t_k,j}, \quad \! \! \!
k\in\emptyset \cup \{1,2,\ldots\}
\end{equation}
where $y(t) = vec \left(y_1(t),\ldots, y_n(t)\right)$, with $y_j(t)\in\mathbb{R}^{n_j},\, j =1, \ldots n$ is the state variable that has known values at instant of time $t_k$, and 
$F_i:\mathbb{R} \times \mathbb{R}^{N} \times \mathbb{R}^{N} \to \mathbb{R}^{N}$ is a given map that characterizes the $i\in\{1, \dots, n\}$ ODE, with $N=n_1+\ldots+n_n$.
Note that here we utilize a unified process to resolve problems involving initial, boundary, or multiple values, so $k \in\emptyset$ refers to the boundary-free functions. 
As an example, suppose that with would like to solve a BVP with two ($n=2$) ODEs, where the first one ($j =1$) starts at $t=t_0$ with some initial condition $y_{0,1}$, and the second one ($j =2$) at $t=t_f$ with a final condition $y_{f,2}$. Then, $y(t) = vec \left(y_1(t), y_2(t)\right)$, and $y_1(t_0)=y_{0,1}$, and $y_2(t_f)=y_{f,2}$.

The goal is to obtain the (approximated) solution $\hat{y}(\cdot)$ of \eqref{DE} along time. For simplicity of presentation and in order to avoid cumbersome notation, at this point, we will consider only one ODE.
The first step is to apply a morph transformation (change of time) so that the problem can be written in the domain of the activation functions to be used in the next steps. More precisely, we make the time transformation $\tau=\tau_0+c( t-t_0)$, with $c>0$.
Next, we impose the solution to be on the form of 
\begin{equation}\label{approximation}   
\hat{y}(\tau)=\theta\left(\tau\right)+\sum\limits_{n=1}^{{k}}{\Omega_{k}\left( \tau \right){(y(\tau_k)-\theta(\tau_k))}}
\end{equation}
where $\theta(\cdot): \mathbb{R}\to \mathbb{R}^N$ indicates a free function, and the scalar function $\Omega_k(\tau)$ is the so-called switching function.
These switching functions are described in \cite{johnston2020fuel} and for $k=1,2$ take the form ${{\Omega }_{1}}\left( \tau \right)=1+\frac{2{{\Delta \tau }^{3}}}{{{\Delta{{\tau }_{f}} }^{3}}}-\frac{3{{ \Delta \tau }^{2}}}{{{ {\Delta{\tau }_{f}}}^{2}}}$ and ${{\Omega }_{2}}\left( \tau \right)=-\frac{2{{\Delta \tau }^{3}}}{{{\Delta{{\tau }_{f}} }^{3}}}+\frac{3{{ \Delta \tau }^{2}}}{{{ {\Delta{\tau }_{f}}}^{2}}}$ with $\Delta \tau=\tau -{{\tau }_{0}}$ and $\Delta \tau_f=\tau_f -{{\tau }_{0}}$.

In this paper, we propose to model the free function $\theta(\tau)$ as a combination of the QNN 
depicted in Fig. \ref{fig:QNN}, where the architecture of each QNN unit is designed by a quantum circuit as illustrated in Fig. \ref{fig:quantum_neural_network}. By considering $L$ qumodes,
the outputs of the QNNs are linearly weighted in accordance with (See Fig. \ref{PIQNN})
 \begin{equation}\label{eq11}
   \theta(\tau)=\sum\limits_{l=1}^{L} \xi_l \sigma(\tau)=\pmb{\sigma}^{T}\left(\tau\right){\pmb{\xi}}
\end{equation}
where $\pmb{\sigma}^{T}\left(\tau\right)=\sigma^T(\tau) \otimes I_{n\times n}$, and the weight $\pmb{\xi}=vec( \xi_1,\ldots, \xi_L )$ is a given parameter.
Now, substituting \eqref{eq11} and \eqref{approximation}  into \eqref{DE}, we then obtain a new set of equations 
\begin{equation}\label{ftild}
F_i(\tau ,\hat{y}(\tau ),\dot{\hat{y}}(\tau ))=\tilde{F}_i(\tau ,{\theta}(\tau ),\dot{{\theta}}(\tau ))=0
\end{equation}
where ${\dot{\theta}(\tau)}=c{{\pmb{\sigma}}^{\prime}}^{ T}(\tau)\pmb{\xi}$ 
with $\pmb{\sigma}^{\prime}(\tau)=\frac{d\pmb{\sigma}(\tau)}{d\tau}$.
Thus, $\tilde F_i(\cdot)$ can be expressed in terms of parameters $\Theta = (\theta, \phi, r,\alpha,\lambda)$ and $\pmb{\xi}$, that is, 
\begin{equation}\label{eq:tilde F}
\tilde F_i(\tau; \Theta, \pmb{\xi})=0.
\end{equation}
The next step consist in defining a loss function to be minimized by updating the parameters in \eqref{eq:tilde F}.  
At this point, we can identify three options to solve \eqref{eq:tilde F}:
\begin{enumerate}
\item Only $\pmb{\xi}$ is updated.
\item Only $\Theta$ is updated.
\item Both parameters $\pmb{\xi}$ and  $\Theta$ are updated.
\end{enumerate}
Option 1) has the advantage of not needing to change the default parameters of the QNN circuit and making the training fast. In this case, a simple approach is to 
discretize $\tau$ into $N$ points, and express the obtained set of differential equations as loss functions at each point as ${{\Xi}}_i^T\left( {{\pmb{\xi}}} \right)=\left[{{{\tilde{F}}}}_i\left( {{\tau}_{0}},{{\pmb{\xi} }} \right), \ldots, {{{\tilde{F}}}}_i\left( {{\tau}_{N}},{{\pmb{\xi} }} \right)\right]$ and obtain the unknown $\pmb{\xi}$, by computing the solution of the augmented loss function $\Xi = \left[\Xi_1^T, \ldots, \Xi_n^T\right]$ associated to the overall system composed by $n$ differential equations.
This can be solved by a numerical minimization scheme, such as for example the iterative least-square method. Through this method, we obtain the iterative procedure 
 ${{\pmb{\zeta} }_{k+1}}={{\pmb{\zeta}}_{k}}+\Delta {{\pmb{\zeta}}_{k}}$, where $\Delta {{\pmb{\zeta}}_{k}}=-{{\left( \mathbb{J}{{\left( {{\pmb{\zeta}}_{k}} \right)}^{T}}\mathbb{J}\left( {{\pmb{\zeta}}_{k}} \right) \right)}^{-1}}\mathbb{J}{{\left( {{\pmb{\zeta}}_{k}} \right)}^{T}}\mathbb{L}({{\pmb{\zeta}}_{k}})$, in which $\mathbb{J}$ is the Jacobian matrix, compiling the partial derivatives of the loss function with respect to each unknown parameter (which here is just $\xi$ and therefore simple to obtain). The iterative procedure continues to be repeated until the convergence criteria is satisfied, meaning that for a predefined tolerance $\bar\varepsilon>0$, we reach to ${{L}_{2}}\left[ \Xi\left( {{\pmb{\zeta} }_{k}} \right) \right]<\bar\varepsilon$.  

Option 2) addresses the case of only updating the QNN parameters $\Theta$. Let $|\psi_\tau\rangle$ be the output state of the circuit given input $D(\tau )|0\rangle$. Then, in this case, the goal is to train the QNN to produce output states whose expectation value for the quadrature operator $\hat \tau$, that is, $\langle\psi_\tau|\hat\tau|\psi_\tau\rangle$ is such that $\tilde F_i(\tau_\ell; \Theta)$ is zero for each sample time $\tau_\ell$.
A simple approach is to use as loss function the mean square error of the obtained values $\tilde F_i(\tau_\ell; \Theta)$ and employ a minimizer (usually based on stochastic gradient descent) to update $\Theta$.

Option 3) makes use of the complete set of parameters. To update them, one need to minimize the loss function, using a backpropagation algorithm which may need the computation of the gradients with respect to the parameters that could be analytical derived or numerically estimated through finite diferences. Another approach is to implement an alternating gradient descent scheme by updating $\Theta$ and $\xi$ in an alternating manner.

The implementation in Fig. \ref{PIQNN} illustrates the proposed integrated quantum-classical approach, aside from what scenario we choose, to solve differential equations and in particular to use it to solve optimal control problems using the indirect method that results in solving a boundary value problem.

\section{Quantum Optimal Control}
The design of quantum optimal controllers is influenced by several factors, including the selection of the cost functional, the formulation of the Pontryagin-Hamiltonian function, and the computational methods used to solve the Pontryagin Minimum Principle (PMP) optimality conditions. In this context, we model the system as an open quantum system, meaning it interacts with its environment. This interaction results in dynamics that are non-unitary yet retain trace-preserving and completely positive properties across all initial conditions. We proceed to outline the optimal control problem by considering a control constrained time-energy minimization problem. To this end, we consider a quantum state transition from the initial state $\rho_i$ to a desired target state $\rho_f$,
governed by the Lindblad master equation.


\subsection{System Description}
Our study takes into account an open quantum system controlled by a time dependent modulation to illustrate the use of physics-informed quantum neural networks for a quantum optimal control problem. We describe open quantum systems by the most general form of Markovian master equations, i.e., the Lindblad form.
The Lindblad master equation for system's density operator $\rho$ can be written as \cite{dehaghani2022quantum}
\begin{equation}\label{lindblad}
 {\dot {\rho }}=-\frac{i}{\hbar}[H\left( u\left( t \right) \right),\rho ]+\sum _{i}^{}\gamma _{i}\left(L_{i}\rho L_{i}^{\dagger }-{\frac {1}{2}}\left\{L_{i}^{\dagger }L_{i},\rho \right\}\right) 
\end{equation}
where $[a , b]= ab-ba$ is the commutator, and ${\displaystyle \{a,b\}=ab+ba}$ is the anticommutator of $a$ and $b$. $H(u(t))$ indicates the system Hamiltonian comprising the drif $H_d$ and control interaction $H_C$ Hamiltonians, i.e., $ H(u(t)) = H_d +H_c u(t)$, where $u(t)$ is the control input.
The \(L_{i}\) symbols denote a group of jump operators that outline the dissipative component of the dynamic processes. The configuration of these jump operators elucidates the environmental impact on the system, which requires accurate determination from the microscopic models that describe the dynamics between the system and its surrounding environment.
Moreover, the \(\gamma_{i} \geq 0\) represent a series of non-negative values referred to as damping rates. 
Corresponding to the dynamics \eqref{lindblad}, we now represent it in terms of the superoperator $\mathcal{L}(\alpha, u(t))$ as following
\[\dot{\Tilde{\rho} }= \mathcal{L}(\alpha, u(t)) \Tilde{\rho} \left( t \right)\]
where $\Tilde{\rho}$ is the state vector, expressed in terms of the elements of density operator ${{\rho }_{ee}}=\left\langle  e \right|\rho \left( t \right)\left| e \right\rangle$, ${{\rho }_{gg}}=\left\langle  g \right|\rho \left( t \right)\left| g \right\rangle$, and ${{\rho }_{ge}}=\left\langle  g \right|\rho \left( t \right)\left| e \right\rangle$ as $\Tilde{\rho}(t)^T=\left( \begin{matrix} 
   {{\rho }_{gg}} & {{\rho }_{ee}} & \operatorname{Re}({{\rho }_{ge}}) & \operatorname{Im}({{\rho }_{ge}})  \\
\end{matrix} \right)$, and $\alpha$ encapsulates the system parameters, i.e., coupling strengths and decay rates.

\subsection{Optimal Control Problems (Indirect Method)}
We deal with a constrained optimal control problem, where the control $u(t)$ belongs to an interval $[\mu^-, \mu^+]$. To this end, we first define a new unconstrained control variable $\nu(t)$. The approach entails replacing the constraint $\mu$ with a smooth and monotonically increasing saturation function defined within the range $u\in [\mu^-, \mu^+]$, such that $u=\phi(\nu)$, where 
\[\phi \left( \nu \right)={{\mu }^{+}}-\frac{\Delta \mu }{1+\exp (\frac{c\nu }{\Delta \mu })},\qquad \Delta \mu ={{\mu }^{+}}-{{\mu }^{-}}\]
for some constant $c>0$ \cite{graichen2010handling}. Optimal control problems typically revolve around a cost function \(J\) , which in its most general form can be expressed as the sum of a terminal cost
and the integral 
over time from \(t_0\) to \(t_f\) 
of the running cost.
Since we introduced a new control variable, we need to also consider an additional term in the cost function. Hence, for our problem, we define the cost function as the following
\begin{align*}
J = \Gamma\, {{t}_{f}}+\eta\int_{{{t}_{0}}}^{t_f}{{{u}^{2}}\left( t \right)\,dt}+\varepsilon\int_{{{t}_{0}}}^{t_f}{{{\nu}^{2}}\left( t \right)\,dt}
\end{align*}
where \(\Gamma\) and \(\eta\) are positive coefficients, with \( t_f \) representing the free final time to be optimized. The second term of \( J \), frequently used in molecular control cost functions, quantifies the energy of the control signal over the interval \([t_0, t_f]\), and the last term is a regularization term defined through a regularization parameter $\varepsilon$. 
\begin{remark}
A low $\varepsilon$ value suggests the new unconstrained control problem approximates the original constrained one. However, as $\varepsilon$ approaches zero, the control variable $\nu$ may rise to infinity, presenting numerical challenges. Therefore, caution is advised in selecting $\varepsilon$ to ensure stability and accuracy in the control outcomes. For more on $\varepsilon$'s impact on control parameters, see \cite{antony2018large}.
\end{remark}

The goal is to minimize
the cost function $J$ subjects to the dynamics $\dot{\Tilde{\rho} }= \mathcal{L}(\alpha, u(t)) \Tilde{\rho} \left( t \right)$.
When using the Pontryagin Minimum Principle (PMP) within the indirect method \cite{ross2015primer}, the Pontryagin Hamiltonian $\mathcal{H}$ for the new unconstrained optimal control problem for all $t\in \left[ {{t}_{0}},{{t}_{f}} \right]$ is formulated by indicating the adjoint variables (costates) $\lambda(t)$ of the system. Hence, 
\begin{equation}\label{pontHam}
\begin{aligned}
&{\mathcal{H}}(\rho,\lambda,u, \nu, \varepsilon, \xi ,t)= \eta {{u}^{2}}\left( t \right)+\varepsilon{{\nu^{2}(t) }}+\lambda^T(t){\mathcal{L}}(\alpha,u(t))\tilde{\rho}(t)\\
&\quad+{\beta(t)\left(u(t)-{{\phi }}\left( \nu  \right) \right)} 
\end{aligned}
\end{equation}
where
$\beta(t)$ is an additional multiplier defined for consideration of equality constraint raised from control constraints. Having defined \eqref{pontHam}, the optimal control law can be derived from the first-order optimality conditions of the PMP, where one obtains 
\begin{equation}\label{u}
    \frac{\partial \mathcal{H}}{\partial u}=\lambda^T(t) \frac{\partial \mathcal{L}}{\partial u}(\alpha,u(t))\tilde{\rho}(t)+2\eta u(t)+\beta(t)=0
\end{equation}
Note that \eqref{u} may result in a transcendental equation, which typically does not yield a closed-form solution for the control. Since a new unconstrained control has been introduced, an additional equation must be incorporated into the optimality condition for the control, which is
\begin{equation}\label{nu}
\frac{\partial {\mathcal{H}}}{\partial \nu}=2\varepsilon\nu-\beta(t)\frac{{{\partial\phi\left( \nu \right) }}}{\partial \nu}=0
\end{equation}
Subsequently, the first-order necessary conditions for the state and costate variables are derived as following
\begin{equation}\label{scos}
    \begin{aligned}
       &\dot{\tilde \rho}=\frac{\partial \mathcal{H}}{\partial \lambda}=\mathcal{L}(\alpha, u(t)) \Tilde{\rho} \left( t \right), \quad \Tilde{\rho}(t_0)=\Tilde{\rho}_i  \\
       &\dot{\lambda}=-\frac{\partial \mathcal{H}}{\partial \tilde \rho}=-\mathcal{L}^T(\alpha, u(t)) \lambda \left( t \right)
    \end{aligned}
\end{equation}
In addition the following transversality conditions are also applicable for our problem;
\begin{equation}\label{tranlam}
\lambda \left( {{t}_{f}} \right)=\lambda_f=\frac{\partial J}{\partial {{\tilde \rho}_{f}}}=\mathbf{0}, \qquad \mathcal{H}\left( {{t}_{f}} \right)=-\frac{\partial J}{\partial {{t}_{f}}}=-\Gamma
\end{equation}
Equations in \eqref{scos}, together with the transversality conditions \eqref{tranlam} applied to the Pontryagin Hamiltonian, and the equality constraint 
\begin{equation}\label{eqc}
    u(t)- \phi(\nu)=0
\end{equation}
constitute a boundary value problem, which we will solve by the techniques explained in the preceding sections.

In accordance with the theory of functional connections in \eqref{approximation}, we approximate the state and costate variables as
\begin{equation*}
\begin{aligned}
    &{{ \tilde{\rho }}}\left( \tau\right)\!=\!\theta_{\tilde{\rho }}(\tau)\!-\!{\Omega}_{1}(\tau)\theta_{\tilde{\rho }}(\tau_0)\!-\!{\Omega}_{2}(\tau)\theta_{\tilde{\rho }}(\tau_f)+{\Omega}_{1}(\tau){ \tilde{\rho }}_i+{\Omega}_{2}(\tau){ \tilde{\rho }}_f\\
    &{{\lambda}}\left( \tau\right)=\theta_{\lambda}(\tau)-{\Omega}_{2}(\tau)\theta_{\lambda}(\tau_f)+{\Omega}_{2}(\tau){ \lambda}_f
\end{aligned}    
\end{equation*} 
By taking the derivatives of the above equation, one obtains
\begin{equation*}
\begin{aligned}
    &\dot{{ \tilde{\rho }}}\left( \tau\right)\!=\!\dot{\theta}_{\tilde{\rho }}(\tau)\!-\!{{\Omega}_{1}^\prime (\tau)}\theta_{\tilde{\rho }}(\tau_0)\!-\!{\Omega}_{2}^\prime(\tau)\theta_{\tilde{\rho }}(\tau_f)+{\Omega}_{1}^\prime(\tau){ \tilde{\rho }}_i+{\Omega}_{2}^\prime(\tau){ \tilde{\rho }}_f\\
    &\dot{{\lambda}}\left( \tau\right)=\dot{\theta}_{\lambda}(\tau)-{\Omega}_{2}^\prime (\tau)\theta_{\lambda}(\tau_f)+{\Omega}_{2}^\prime (\tau){ \lambda}_f
\end{aligned}    
\end{equation*}
In addition, for control variables and also the equality constraint multiplier, we consider the functional approximation as a free function only, i.e., 
\begin{equation*}
 u( \tau)=\theta_u(\tau), \quad \nu( \tau)=\theta_\nu(\tau), \quad \beta( \tau)=\theta_\beta(\tau)
\end{equation*}
Now, we are in the position to define an augmented loss function to be minimized. Note that the setup indicated in Fig. \ref{PIQNN} is for minimization of only one loss function. To solve the problem for an augmented loss function, we consider an augmented network that minimizes all elements of the augmented loss simultaneously which are indicated as
\begin{equation*}
\begin{aligned}
&\Xi_{\Tilde{\rho}}= \dot { \tilde{\rho }}-{\mathcal{L}}(\alpha, u(t)) \Tilde{\rho}(t) \\
&\Xi_{\lambda}=\dot{\lambda}+\mathcal{L}^T(\alpha, u(t)) \lambda \left( t \right)\\
&\Xi_u=\lambda^T(t) \frac{\partial \mathcal{L}}{\partial u}(\alpha,u(t))\tilde{\rho}(t)+2\eta u(t)+\beta(t)\\
&\Xi_\nu=2\varepsilon\nu-\beta(t)\frac{{{\partial\phi\left( \nu \right) }}}{\partial \nu}\\
&\Xi_\phi=u(t)-\phi(\nu)\\
&\Xi_{{\mathcal{H}}}={\mathcal{H}}\left( t_f \right)+\Gamma\\
\end{aligned}
\end{equation*}
leading to an augmented form of the loss function as
\begin{equation}\label{51}
    \mathbb{L}= \left[\begin{matrix}
   \Xi_{\Tilde{\rho}} &  \Xi_{\lambda} & \Xi_u & \Xi_\nu & \Xi_\phi & \Xi_{{\mathcal{H}}}  \\
\end{matrix} \right]^T
\end{equation}
which will be minimized during the procedure.

\section{Simulations and Results}
\begin{figure*}[t]
  \includegraphics[width=\textwidth]{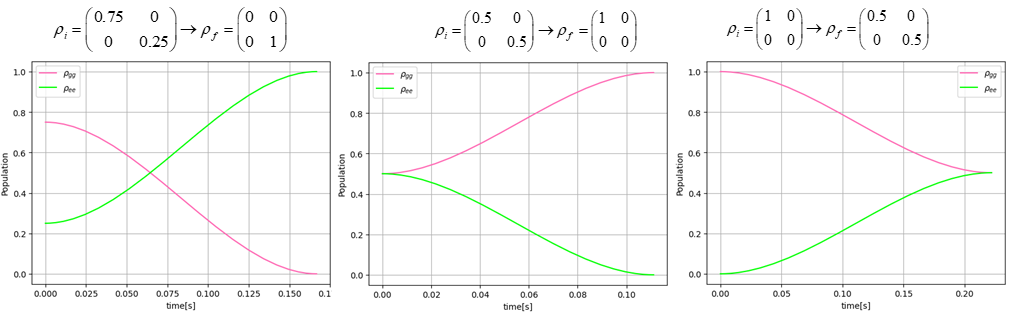}
\caption{The plots illustrate the time evolution of populations in a two-level quantum system from different initial states to specific target states, highlighting the dynamics of the ground and excited state probabilities over time. Each plot shows a distinct trajectory of population transfer, reflecting various quantum processes such as relaxation and population inversion.}
  \label{ST}
\end{figure*}

We consider an open two-level quantum system (a qubit) controlled by a time dependent modulation as a principle example to illustrate the use of physics-informed quantum neural networks for a quantum optimal control problem. In this regard, quantum transitions can be written in terms of the operators
${{\sigma }_{ab }}=\left| a  \right\rangle \left\langle  b  \right|$ where $\left| a\right\rangle, \left| b  \right\rangle= \left| e  \right\rangle, \left| g  \right\rangle$ being $\left| e  \right\rangle$ and $\left| g  \right\rangle$ the excited and ground state, respectively. 
Consider the quantum system Hamiltonian 
with a phase damping control $u(t)$ where the drift term is defined through the system parameters $\omega_x$ and $\omega_z$ as
\[{{H}_{d}}={{\omega }_{x}}\left( {{\sigma }_{eg}}+{{\sigma }_{ge}} \right)+{{\omega }_{z}}\left( {{\sigma }_{ee}}-{{\sigma }_{gg}} \right)\] 
and the interaction control Hamiltonian is $H_c={\sigma }_{ee}$.
We consider two jump operators, i.e., for absorption $L_1= \sigma_{eg}$ with decay rate $\gamma_{eg}$ and emission $L_2= \sigma_{ge}$ with decay rate $\gamma_{ge}$. Hence, the dynamics in terms of the superoperator $\mathcal{L}(\alpha, u(t))$ is given by \cite{norambuena2024physics} 
\[\dot{x}=\left( \begin{matrix}
   -{{\gamma }_{eg}} & {{\gamma }_{ge}} & 0 & -2{{\omega }_{x}}  \\
   {{\gamma }_{eg}} & -{{\gamma }_{ge}} & 0 & 2{{\omega }_{x}}  \\
   0 & 0 & \frac{-\left( {{\gamma }_{ge}}+{{\gamma }_{eg}} \right)}{2} & -\left( 2{{\omega }_{z}}+u(t) \right)  \\
   {{\omega }_{x}} & -{{\omega }_{x}} & 2{{\omega }_{z}}+u(t) & \frac{-\left( {{\gamma }_{ge}}+{{\gamma }_{eg}} \right)}{2}  \\
\end{matrix} \right)x\left( t \right)\]
We consider the parameters as follows: ${\gamma }_{eg}=0.1$, ${\gamma }_{ge}=0.3$, ${{\omega }_{x}} =1$, ${{\omega }_{z}} =2$. In Fig. \ref{ST}, we have shown different state transitions processes.  The time-dependent behavior of the ground state population and the excited state population is depicted for three different initial state configurations transitioning towards their respective target states. The plots succinctly capture the essence of quantum state transitions, ranging from complete population transfer to the achievement of a balanced superposition, showcasing the intricate nature of quantum dynamics.

\begin{figure*}[t]
  \includegraphics[width=\textwidth]{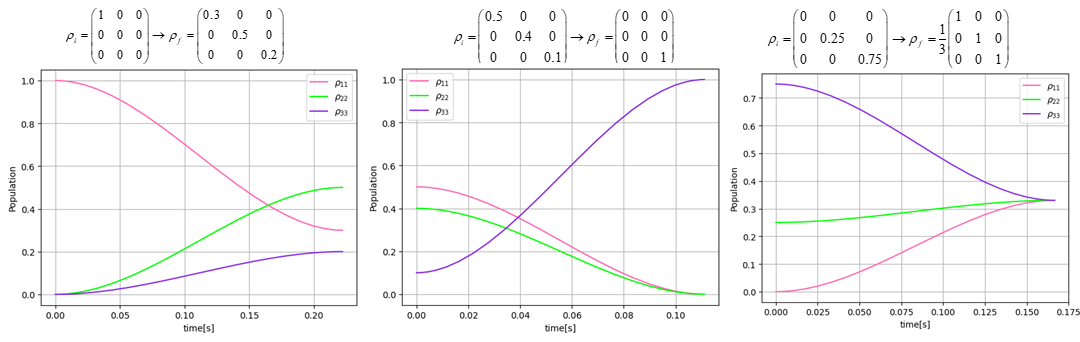}
\caption{The figure displays the evolution of state populations in a three-level quantum system across three distinct scenarios. Each plot tracks the probabilities of the system over time, beginning from different initial mixed states and converging towards predetermined target states. The dynamics reveal transitions such as population inversion, decay to the ground state, and equilibration to a specific superposition, demonstrating the versatile control achievable in quantum state engineering.}
  \label{ST3}
\end{figure*}

To show the scalability of our method, we also demonstrate quantum state transition for a three-level system. Three-level quantum systems, also known as qutrits in the context of quantum computing, are deemed to be significant in various areas of quantum research due to their potential to represent more complex states than two-level systems and their applications in quantum information processing, quantum simulations, and precise control in quantum optics and spectroscopy \cite{petiziol2020optimized,dehaghani2022quantum, sugny2008optimal,d2024optimal}. The Hamiltonian for the three-level system under study is given by:
\[
{H} = \delta\sigma_{22} + \Delta_{1}\sigma_{33} + \left( \frac{u_p(t)}{2}\sigma_{13} + \frac{u_s(t)}{2}\sigma_{23} + h.c. \right),
\]
where the parameters are defined as follows: \(\delta\)=0.1 is the two-photon detuning, \(\Delta_{1}\) is the one-photon detuning, calculated as \(\Delta_{1} = \omega_0 - \omega_p - \omega_s\), where \(\omega_0\) is the transition frequency between the ground state and the first excited state, \(u_p(t)\) and \(u_s(t)\) are the Rabi frequencies for the pump and Stokes fields, respectively.
and \(\sigma_{ij}\) are the transition operators between the states \(i\) and \(j\).
calculated as before. The dynamics of the system are influenced by the interaction terms \(\sigma_{13}\) and \(\sigma_{23}\), which couple the ground state to the two excited states via the external fields characterized by their respective Rabi frequencies. These interactions are crucial for the control schemes aimed at achieving state transitions or implementing quantum gates within this three-level system. The results can be seen in Fig. \ref{ST3}. 
The depicted figure provides a compelling visualization of the temporal dynamics of state populations within a three-level quantum system under controlled transitions. For each subplot, we initiate the system in a distinct mixed state, observing its evolution towards a designated target state. These evolutions underscore the nuanced manipulation of quantum states, which is vital for applications in quantum computing, communication, and sensing.

\section{Conclusion}
In conclusion, our study presents a hybrid quantum-classical architecture that innovatively integrates the principles of quantum mechanics with classical neural network techniques. By employing a dynamic quantum circuit and leveraging Physics-informed neural networks, we have successfully addressed and solved quantum optimal control problems. The architecture's adeptness in managing complex state transitions highlights its versatility and significant potential across various quantum computing applications. Our approach paves the way for novel methodologies in quantum control, with potential applications to quantum technology applications in computing, secure communications, and precision metrology.

\section*{Acknowledgment}
The authors acknowledge the support of FCT for the grant 2021.07608.BD, the ARISE Associated Laboratory, Ref. LA/P/0112/2020, and the R$\&$D Unit SYSTEC-Base, Ref. UIDB/00147/2020, and Programmatic, Ref. UIDP/00147/2020 funds, and also the support of projects SNAP, Ref. NORTE-01-0145-FEDER-000085, and  RELIABLE (PTDC/EEI-AUT/3522/2020) funded by national funds through FCT/MCTES. The work has been done in the honor and memory of Professor Fernando Lobo Pereira.

\bibliographystyle{IEEEtran}
\bibliography{ref}

\end{document}